# Ferroelectric metal-oxide-semiconductor capacitors using ultrathin single crystalline SrZr$_x$Ti$_{1-x}$O$_3$

Reza M. Moghadam[1], Zhiyong Xiao[2], Kamyar Ahmadi-Majlan[1], Everett D. Grimley[3], Mark Bowden[4], Phuong-Vu Ong[5], Scott A. Chambers[5], James M. Lebeau[3], Xia Hong[2], Peter V. Sushko[5] and Joseph H. Ngai[1*]

[1]Department of Physics, University of Texas Arlington, Arlington, TX 76019, USA
[2]Department of Physics and Astronomy, University of Nebraska Lincoln, Lincoln, NE 68588, USA
[3]Department of Materials Science and Engineering, North Carolina State University, Raleigh, NC 27695, USA
[4]Environmental Molecular Sciences Laboratory, Earth & Biological Sciences Directorate, Pacific Northwest National Laboratory, Richland, WA 99352, USA
[5]Physical Sciences Division, Physical & Computational Sciences Directorate, Pacific Northwest National Laboratory, Richland, WA 99352, USA
*email: jngai@uta.edu

The epitaxial growth of multifunctional oxides on semiconductors has opened a pathway to introduce new functionalities to semiconductor device technologies. In particular, ferroelectric materials integrated on semiconductors could lead to low-power field-effect devices that can be used for logic and memory. Essential to realizing such field-effect devices is the development of ferroelectric metal-oxide-semiconductor (MOS) capacitors, in which the polarization of a ferroelectric gate is coupled to the surface potential of a semiconducting channel. Here we demonstrate that ferroelectric MOS capacitors can be realized using single crystalline SrZr$_x$Ti$_{1-x}$O$_3$ (x = 0.7) that has been epitaxially grown on Ge. We find that the ferroelectric properties of SrZr$_x$Ti$_{1-x}$O$_3$ are exceptionally robust, as gate layers as thin as 5 nm corresponding to an equivalent-oxide-thickness of just 1.0 nm exhibit a ~ 2 V hysteretic window in the capacitance-voltage characteristics. The development of ferroelectric MOS capacitors with nanoscale gate thicknesses opens new vistas for nanoelectronic devices.

Ferroelectrics integrated on semiconductors have long been proposed to serve as a materials platform for a variety of electronic device technologies [1] [2]. These proposed technologies envision utilizing the ferroelectric as a field-effect gate material in which its re-orientable polarization is coupled to itinerant carriers in a semiconducting channel. For applications in sensing, ferroelectric gate materials could lead to smart transistors that are sensitive to temperature and pressure [3]. In regards to computing, stabilization of the negative capacitance of ferroelectric gate materials could lead to field-effect transistors that require very little power to operate [4]. Ferroelectric gate materials could also lead to field-effect devices for non-volatile logic and memory due to the hysteretic remnant polarization that persists after power is removed [5]. Such hysteretic devices could also be exploited in computing based on emerging neuromorphic architectures [6]. In short, a materials platform that combines ferroelectrics and semiconductors could lead to field-effect devices that far surpass the performance and functionality of present devices.

In this regard, advancements in thin film growth have enabled single crystalline ferroelectrics to be epitaxially integrated on semiconductors [7], such as $Pb(Zr_{0.2}Ti_{0.8})O_3$ on Si [8], $BaTiO_3/Ba_xSr_{1-x}TiO_3$ on Si [9] [10], $SrTiO_3$ on Si [11], $BaTiO_3$ on GaAs [12] and $BaTiO_3$ on Ge [13] [14]. The growth of crystalline ferroelectrics on semiconductors offers a pathway to achieve material properties that are ideal for device applications, such as mono-domain states, and improved interfacial structure with the semiconductor. In addition, the properties of crystalline oxide thin films can be controlled and enhanced through the effects of epitaxial strain [15].

Despite such progress, the development of scalable, metal-oxide-semiconductor (MOS) capacitors using single crystalline ferroelectric gate materials remains elusive. MOS capacitors enable the surface potential of a semiconductor to be modulated through an applied bias, and thus underpin the functionality of field-effect devices. Modulation of the surface potential gives rise to a change in the capacitance of the semiconducting electrode, which is manifested as dispersion in the capacitance-voltage (C-V) characteristics of a MOS capacitor. When the re-orientable polarization of a ferroelectric gate material is coupled to the surface potential, hysteresis in the C-V characteristics typically emerges [16]. Realizing ferroelectric MOS capacitors is challenging for several reasons. First, coupling of the polarization of a ferroelectric to the surface potential of a semiconductor requires continuity in the electric displacement between materials that differ in elemental composition and bonding. Dangling bonds at interfaces between a ferroelectric and a semiconductor can trap charges that electrically screen the polarization. Second, a type-I arrangement in which the conduction (valence) band of the ferroelectric is above (below) the conduction (valence) band of the semiconductor is needed to enable ferroelectric MOS structures to function as capacitors [17]. Unfortunately, ferroelectrics such as $BaTiO_3$ exhibit a type-II band alignment with respect to Si, Ge and GaAs, in which the conduction band of the oxide is below that of the semiconductor [18] [19]. Third, for applications in computing, the continued lateral scaling of field-effect transistors places constraints on the thickness of gate materials that can be used in practical devices [20] [21]. For example, present state-of-the-art transistors utilize gate oxides that are only a few nanometers thick [22]. However, ferroelectricity is generally weakened at such thicknesses because of depolarization fields, which are particularly strong for ferroelectrics grown on semiconductors due to longer screening lengths in the latter [23]. Due to the above challenges, prior reports of single crystalline ferroelectric MOS capacitors have typically utilized thick ($\geq$ 90 nm) ferroelectric layers [24] [25].

Here we present ferroelectric MOS capacitors comprised of single crystalline $SrZr_xTi_{1-x}O_3$ $x = 0.7$ (SZTO) that has been grown on Ge, a semiconductor that is of interest for next-generation field-effect transistors due to its high carrier mobility. The ferroelectric properties of SZTO have been characterized using piezo-response force microscopy (PFM), polarization-voltage PUND, and C-V measurements. Analysis of high angle annular dark-field (HAADF) images obtained through scanning transmission electron microscopy (STEM) reveals evidence for relaxor behavior, namely, regions that exhibit non-centrosymmetric displacements of Ti/Zr cations, consistent with the presence of polar nano-regions (PNRs) [26] [27]. *Ab initio* density functional theory (DFT) calculations suggest that non-centrosymmetric displacements of Ti cations substituted in a sublattice dominated by larger Zr cations play a key role in the formation of PNRs. We find that the ferroelectricity in SZTO is exceptionally robust, as 5 nm thick films corresponding to an equivalent-oxide-thickness (EOT) of just 1.0 nm exhibit a ~ 2 V wide hysteretic window in the C-V characteristics, opening a pathway to realize highly scalable ferroelectric field-effect devices.

Epitaxial SZTO films are grown on p-type Ge using oxide molecular beam epitaxy (MBE). As grown films are electrically leaky, which is attributed to residual oxygen vacancies formed in the relatively low oxygen background pressure of the ultra-high vacuum MBE chamber. To minimize the presence of residual vacancies, the SZTO films are annealed *ex situ* in oxygen after growth. We present electrical characterization of annealed, 15 and 5 nm thick SZTO films on Ge below. Aside from minimizing residual vacancies, annealing can create a thin layer of GeO$_x$ at the interface as shown in Supplementary Fig. 1 and Fig. 3d for the 15 and 5 nm thick films, respectively. The annealed films are compressively strained ($a$ = 4.03 Å and $c$ = 4.07 Å) relative to bulk SZTO of the same $x$ = 0.7 Zr content ($a$ = 4.048 Å and $c$ = 4.052 Å) [28], as shown in Supplementary Fig. 2.

Stable ferroelectric domains can be written on SZTO using PFM. Figures 1a and 1b show spatial mappings of phase and amplitude response for a domain structure written on a 15 nm thick film. Here, a conductive atomic-force-microscope (AFM) tip held at a bias of -10 V is used to write a rectangle, followed by an adjacent rectangle written at a tip bias of +10 V. The topography, shown in the inset of Fig. 1e, indicates that no irreversible electrochemical reactions occur on the surface due to poling. Figures 1c and 1d show the phase and amplitude responses, respectively, of the same region 48 hours after poling. The amplitude response is stable for both polarization orientations for at least 48 hours, which was the duration of measurement, as summarized in Fig. 1e. An initial drop in the amplitude response immediately after writing is attributed to a surface charging effect. The persistence of spatially poled regions for extended periods of time in the absence of electrochemical changes on the surface is consistent with the creation of ferroelectric domains. Ferroelectric switching is also manifested in piezo-response spectroscopy measurements, in which the tip is held in one location and the phase and amplitude response are measured as a function of applied bias. Figures 1f and 1g show the amplitude and phase responses of the tip, revealing ferroelectric butterfly-like hysteresis, and 180° phase shift, respectively. The PFM measurements also reveal that as-annealed films are preferentially poled in the downwards direction, i.e. with a polarization oriented towards the Ge electrode. The downwards polarization is so strong in some nanoscale regions that a -10 V tip bias is insufficient to switch the polarization upwards. Regions that exhibit asymmetric coercive voltages that favor a downwards polarization are also observed in piezo-response spectroscopy measurements (not shown).

Phase and/or amplitude PFM contrast have been observed in single crystalline $A$TiO$_3$ ($A$ = Sr, Ba) films grown on semiconductors, such as BaTiO$_3$ on Si [9] [10], SrTiO$_3$ on Si [11], BaTiO$_3$ on GaAs [12], and BaTiO$_3$ on Ge [14]. However, $A$TiO$_3$ has a type-II band alignment with respect to Si, Ge, and GaAs [19] [18] [13]. In contrast, SZTO exhibits a type-I band alignment with respect to Ge, thus robust MOS capacitors can be realized [29] [30].

Capacitance-voltage measurements demonstrate that the ferroelectric polarization of SZTO is coupled to the surface potential of Ge, which is essential for device applications. Figure 2a shows C-V characteristics of 15 nm thick SZTO taken between ± 3 V, ± 4 V and ± 5 V. Hysteresis is observed, with an orientation with respect to the direction of the sweeping voltage that is consistent with switching of ferroelectric dipoles, as opposed to hysteresis induced by interfacial trap states. Mobile ions also cannot account for the hysteresis observed in our capacitors, since a positive flat-band voltage is achieved, and the hysteresis does not diminish by increasing the rate at which the voltage is swept between positive and negative values, as shown in Supplementary Fig. 3 [5]. The enhancement in hysteresis with magnitude of applied voltage is attributed to an increase in remnant polarization with electric field, discussed below.

The type-I band offset also enables the hysteretic, or remnant component of polarization to be estimated. Measurements are performed using the PUND technique [31] [32] to isolate the hysteretic component of current due to ferroelectric switching from leakage currents, the latter of which are sizeable due to the thinness of the SZTO films. Figure 2b shows the current (black) measured in response to the PUND waveform (grey), consisting of switching P, N and non-switching U, D voltage ramps applied for positive and negative polarities to a Ni electrode. The magnitude of leakage current is larger for positive P, U voltage ramps in comparison to negative N, D voltage ramps, due to a difference in barrier height for charge injection, as illustrated by the band diagrams shown as insets in Fig. 2b. Ferroelectric switching is clearly resolved, as excess current on the rising (falling) side of the P (N) voltage ramp is measured relative to the rising (falling) side of the U (D) ramp as shown in Fig. 2c. The excess switching current is distributed over a broad voltage range in Fig. 2c, indicating the absence of a well-defined coercive voltage, which is likely due to some nanoscale variation in electronic properties. Integration of the excess currents in P and N voltage ramps results in the two-respective polarization half-loops shown in Fig. 2d. Note the different scales used for each polarization half-loop, as the amount of charge switched for positive polarity P ramp is larger than the charge for negative polarity N ramp. The difference in switching charge between P and N polarities could be due to several factors. The leakage currents are very large in comparison to the switching currents, rendering precise measurement of the latter difficult. Leakage currents that are a function of the polarization could also give rise to the observed difference [13]. Thus, we estimate the switching charge ranges from ~ 3 to 5.5 µC/cm$^2$, which corresponds to a remnant polarization of ~ 1.5 to 2.8 µC/cm$^2$, for voltage ramps of ± 3 V on a 15 nm thick SZTO film. We find that the remnant polarization can be enhanced by increasing the magnitude of the voltage ramps, as shown in Supplementary Fig. 4 for PUND measurements taken between ± 4 V. The enhancement of remnant polarization with increasing magnitude of applied voltage, i.e. electric field, indicates that the polarization of SZTO has not reached saturation [5] [24].

Yet more remarkable, we find that the ferroelectricity in SZTO grown on Ge is robust for films as thin as 5 nm. Figure 3a shows the C-V characteristics of 5 nm thick film taken between ± 3 V, which results in a 2 V wide hysteretic window. Ferroelectric domains can be written on the 5 nm thick SZTO films using PFM, as shown in the inset of Fig. 3a, and piezo-response spectroscopy also confirms butterfly-like hysteresis and 180° phase-shift in the amplitude and phase response, shown in Fig. 3b and 3c, respectively. The annealed 5 nm thick SZTO films also exhibit relatively abrupt interfaces with Ge, as shown in the HAADF STEM images in Fig. 3d. An average ~ 0.6 nm thick layer of interfacial GeO$_x$ is achieved with regions that even exhibit no discernible GeO$_x$ (top panel) using our basic annealing set-up. In principle, more advanced techniques such as rapid thermal annealing should enable better control of the oxidation process to further minimize GeO$_x$. The 5 nm SZTO yields an EOT of 1.0 nm, which is approaching the EOT of gate oxides used in present conventional MOS transistors. The large hysteretic window in C-V achieved with an EOT of 1.0 nm opens a pathway to realize highly scalable ferroelectric field-effect devices.

STEM imaging reveals evidence of non-uniform atom column displacements in the films which are consistent with PNRs associated with relaxor materials. Spatial regions that exhibit non-centrosymmetric displacements of B-site cations relative to the A-site sub-lattice are a signature of PNRs in ABO$_3$ structured perovskites. To look for such displacements, mappings of the deviations of the Zr/Ti B-site atom columns from the ideal center of the perovskite cells bounded by their four neighboring A-site Sr atom columns are performed on HAADF images of a 5 nm thick film, as shown in Fig. 3e. The directions and lengths of the arrows superimposed on the HAADF image represent the directions and magnitudes, respectively, of the non-centrosymmetric Zr/Ti displacements. Consistent with the presence of PNRs, spatial correlation between the magnitudes and directions of the non-centrosymmetric displacements is observed, as regions that contain larger-than-average displacements are aligned in similar directions. In contrast, the magnitudes and the directions of displacements are typically random in non-polar structures [33]. Mapping the deviation of a dumbbell from the center of its four neighboring dumbbells in the Ge substrate reveals typically smaller displacements that are not aligned, as shown in Fig. 3e. In general, we find that regions exhibiting non-centrosymmetric Zr/Ti displacements vary in size, shape, and direction of displacement. For 5 nm thick films, the regions containing ordered displacements are found to align with preference towards the $+c$ and $-c$ directions of the film. In standard $ABO_3$ perovskite structured relaxors, the origin of PNRs is linked to disorder in the composition of the B-site cation. In this regard, SZTO is similar to $BaZr_xTi_{1-x}O_3$, which is a relaxor that has been extensively characterized [34]. Unlike $BaZr_xTi_{1-x}O_3$, however, relaxor behavior has yet to be reported in bulk SZTO [35].

To gain insight on how PNRs form in strained SZTO thin films, we performed *ab initio* DFT modeling of this system. We first seek to understand the behavior of Ti cations substituted into a B-site sublattice dominated by Zr, as shown in Fig. 4a. In pure $SrZrO_3$ (SZO) that is compressively strained to the lattice constant of Ge, all in-plane and out-of-plane Zr–O distances are 2.05 and 2.12 Å (see Fig. 4b), respectively, indicating that bulk SZO does not exhibit lattice polarization even under compressive strain. In contrast, upon substituting a Ti atom into a Zr site, the out-of-plane $Ti_{Zr}$–O bond lengths split into short (1.92 Å) and long (2.16 Å) ones, as shown in Fig. 4b (hollow squares). This effect is attributed to the smaller ionic radius of $Ti^{4+}$ (74.5 pm) in comparison to $Zr^{4+}$ (86 pm), which leads to a double-well potential energy surface for $Ti^{4+}$ ions and, in turn, results in effectively 5-fold coordination of the $Ti_{Zr}$ site. Importantly, the distortion of the Ti local environment induces disproportionation of Zr–O bond lengths into short (2.06 Å) and long (2.25 Å) ones at the neighboring Zr sites along the *c*-axis, (hollow squares in Fig. 4b), which indicates that not only are Ti-centered octahedra polarized themselves, but they also polarize neighboring Zr-centered octahedra.

To understand the cumulative effect of such displacements in a film which has random placement of Ti amongst a Zr sublattice, we consider 4 types of configurations of Ti cations, namely, column, screw, pairs and plane with an overall Ti concentration of 25%, as shown in

Fig. 4c. The plane configuration was found to be energetically more favorable than the column (+0.1 eV) and the screw (+0.35 eV) configurations. However, given the small Ti to Zr ratio and the substrate temperatures used during film growth, the predominance of planar structures is unlikely. The remaining configurations are within 0.25 eV (0.05 eV per Ti) from each other, and we propose that all of them are equally likely to be realized. A periodic, 2×2 lateral supercell slab is used to model the SZTO film on a SrTiO$_3$ substrate, as shown in Fig. 4d, in which the in-plane lattice parameters are fixed to the experimental value of $a = b = 4.03$ Å, and the out-of-plane parameter is $c = 50.0$ Å. We choose SrTiO$_3$ (STO) as a substrate to bypass the complexities associated with modelling a SZTO/Ge interface with a partially oxidized GeO$_x$ layer. In Fig. 4e we plot the calculated bond lengths ΔZ along the *c*-axis between B-site (Ti,Zr) cations and oxygen anions located in adjacent planes. Each plane within the periodic slab is numbered as shown in Fig. 4c. The colors indicate the different sets of the lateral fractional coordinates: black (0,0), red (1/2,0), green (0,1/2), blue (1/2,1/2) within the 2×2 lateral supercell. For example, black (0,0) in the column configuration would refer to the continuous chain of Ti cations along the *c*-axis shown in Fig. 4c. If all Ti cations are situated in a single column along the *c*-axis, we find that a continuous sequence of short-long-short-long-etc., Ti–O bonds forms. If Ti cations are in different columns in a screw or pairs configuration, short-long Ti/Zr–O bond length sequences also form, though the difference in length between short and long bonds is smaller. Finally, for the plane configuration, long-long-short-short-long-long bond sequences appear that are locally symmetric with respect to the Ti plane.

The sequences of Ti/Zr–O bond lengths give rise to a dipole moment within each BO$_2$ plane. Figure 4f summarizes the out-of-plane contributions to polarization from each BO$_2$ plane for the 4 configurations (see Supplementary Information for details on calculation). We note that in-plane components of polarization are also produced (not shown). The column configuration gives rise to a net polarization that is long-range, i.e. extends throughout the thickness of the film, while screw, pairs and plane configurations lead to polarizations that are shorter in range and smaller in magnitude. These results also yield insight as to the predominance of a downward oriented polarization. The existence of the surface/vacuum interface gives rise to an asymmetry in the structure of the film and, accordingly, lattice polarization. Since Ti–O bonds are shorter than Zr–O bonds, Ti atoms in the top plane displace down with respect to the average Zr position in the same plane, and thus favors the short-long-short-… Ti/Zr–O bond pattern that corresponds to a downward polarization. Aside from the dominant downward polarization at the surface, we find that the bulk corresponds to virtually zero net polarization because the statistical probability of distorted Ti/Zr-centered octahedra to contribute to upwards and downwards polarizations are the same everywhere except at the surface and the interface with the substrate. We note that our calculations do not consider defects, vacancies and variations in strain, all of which can affect the nucleation of PNRs. Furthermore, our calculations do not account for screening of polarization via charges in the substrate or adsorbates on the surface, or applied electric fields, all of which can affect the size and alignment of PNRs. Nevertheless, our calculations elucidate a defining feature of relaxors, namely, a ground state characterized by local dipoles that spatially vary in a random manner.

In summary, we have realized ferroelectric MOS capacitors exhibiting ~ 2 V wide hysteretic C-V characteristics with a corresponding EOT of just 1.0 nm using 5 nm thick, single crystalline SZTO. Through the epitaxial growth of SZTO on Ge, we have achieved continuity in the electric displacement, a type-I band offset, and robust ferroelectricity at nanoscale film thicknesses. The emergence of ferroelectricity in thin films of SZTO on Ge demonstrates how reduced dimensionality, strain and applied electric fields can be exploited to realize technologically important material behavior in multifunctional oxides. The realization of a materials platform that combines nanoscale ferroelectrics with semiconductors opens new vistas for nanoelectronic devices.

**Acknowledgements**

This work was supported by National Science Foundation (NSF) under award No. DMR-1508530. The work performed at the University of Nebraska was supported by DOE (BES) Award DE-SC0016153 funded by the U.S. Department of Energy, Office of Basic Energy Sciences for the scanning probe studies. The work performed at North Carolina State University was supported by the NSF under award No. DMR-1350273. E.D.G. acknowledges support for this work through a National Science Foundation Graduate Research Fellowship (Grant DGE-1252376). This work was performed in part at the Analytical Instrumentation Facility (AIF) at North Carolina State University, which is supported by the State of North Carolina and the National Science Foundation (award number ECCS-1542015). The AIF is a member of the North Carolina Research Triangle Nanotechnology Network (RTNN), a site in the National Nanotechnology Coordinated Infrastructure (NNCI).  DFT, and x-ray diffraction performed at Pacific Northwest National Laboratory was supported by the U.S. Department of Energy, Office of Science, Division of Materials Sciences and Engineering under Award #10122. The PNNL work was performed in the Environmental Molecular Sciences Laboratory, a national scientific user facility sponsored by the Department of Energy's Office of Biological and Environmental Research and located at PNNL.

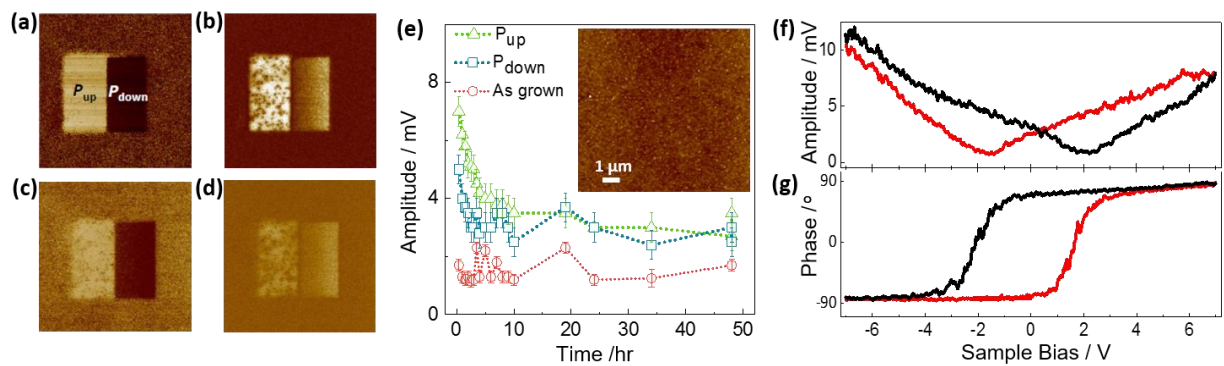

**Figure 1**. Piezo-response force microscopy of 15 nm thick SZTO film on Ge. (a) Phase and (b) amplitude response of domain structures 2.5 hours after writing. Lateral scale is the same as shown in the inset of (e). (c) Phase and (d), amplitude response of the same domain structures 48 hours after writing. (e) Amplitude response as a function of time for $P_{UP}$ and $P_{DOWN}$ domain structures, and as-annealed parts of film surface. Inset shows the topography of the film after domain writing, indicating no tip induced electrochemical changes on the surface. (f) Amplitude and (g), phase piezo-response taken in spectroscopy mode, showing butterfly-like hysteresis and 180° phase change, respectively.

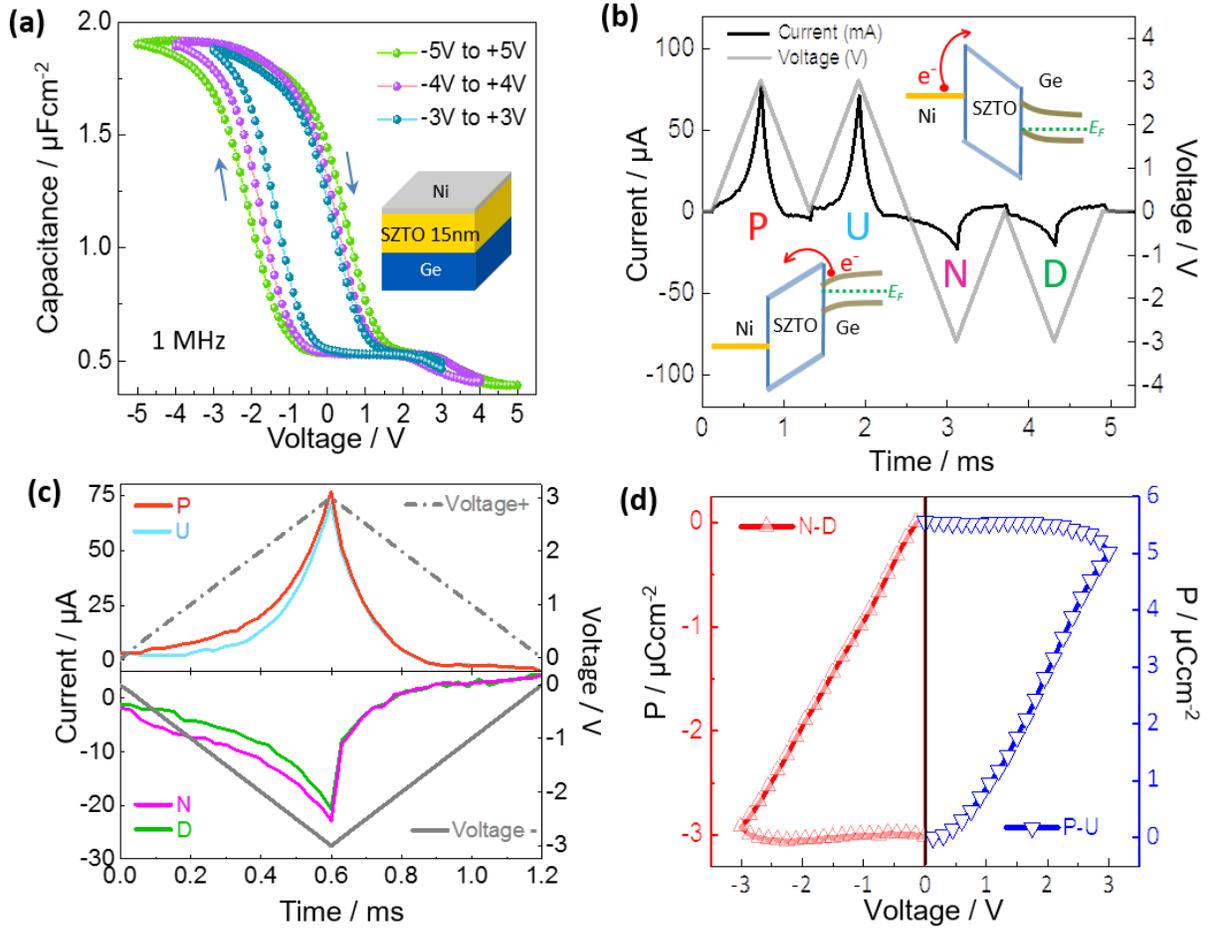

**Figure 2**. *C-V* and *P-V* characterization for 15 nm thick SZTO film on Ge. (a) *C-V* measurements showing hysteresis obtained for ± 3 V, ± 4 V and ± 5 V sweeps. (b) Current (black) measured through the Ni-SZTO-Ge stack in response to the PUND waveform (grey) applied to the Ni electrode. Bottom left (top right) inset illustrates the band diagram of the Ni-SZTO-Ge stack under positive (negative) bias. (c) Current measured from each voltage ramp of the PUND waveform plotted independently on the same time scale to enable comparison. Switching of the hysteretic component of ferroelectric polarization is observed as excess current is measured on the rising (falling) side of the P (N) voltage ramp relative to the rising (falling) side of the U (D) ramp, as shown in the upper (lower) panel. (d) *P-V* half-loops obtained by integrating switching currents from the PUND measurements taken using ± 3 V voltage ramps.

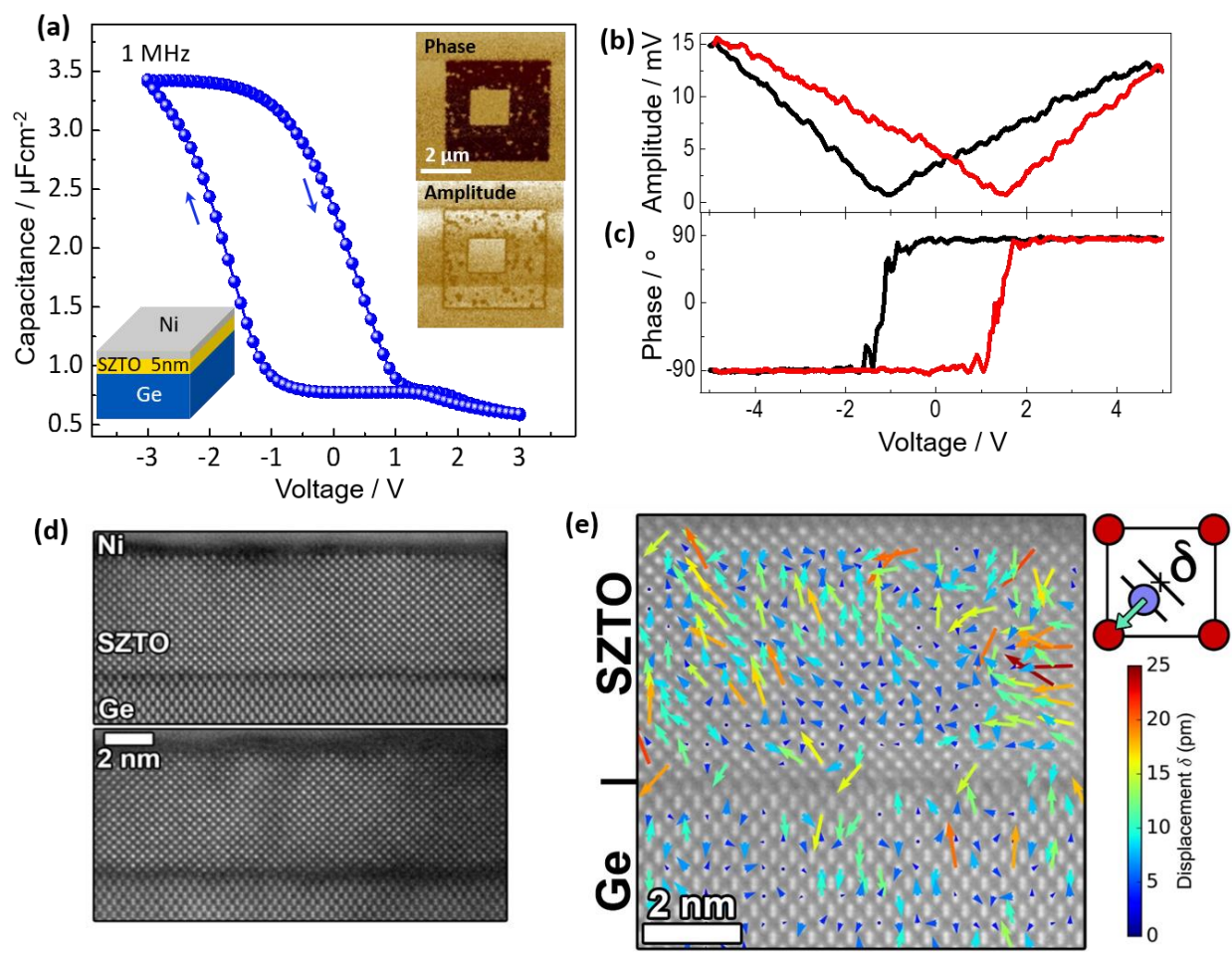

**Figure 3.** Electrical and structural characterization of 5 nm SZTO film on Ge. (a) *C-V* measurement of 5 nm thick SZTO film on Ge. Inset shows phase and amplitude contrast of domain structures written using PFM. (b) Amplitude and (c) phase piezo-response spectroscopy measurements of 5 nm thick SZTO on Ge. (d) HAADF STEM images taken on annealed 5 nm thick SZTO on Ge, showing relatively abrupt interfaces. Both images are from the same lamella, and show the variation of $GeO_x$ formation at the interface, from little-to-no $GeO_x$ in the top panel to containing regions around 0.5 nm thick as in portions of the bottom panel. On average, there is ~ 0.6 nm thick interfacial $GeO_x$ between the 5 nm thick SZTO film and Ge. (e) Magnitude and direction of non-centrosymmetric Zr/Ti column displacements δ superimposed on HAADF image of annealed 5 nm thick SZTO film on Ge. Top right schematic shows how δ is measured with respect to *A*-site and *B*-site cations. The color scale also indicates the magnitude of displacement, which was limited to 20 pm to emphasize the spread in color over the relevant distances.

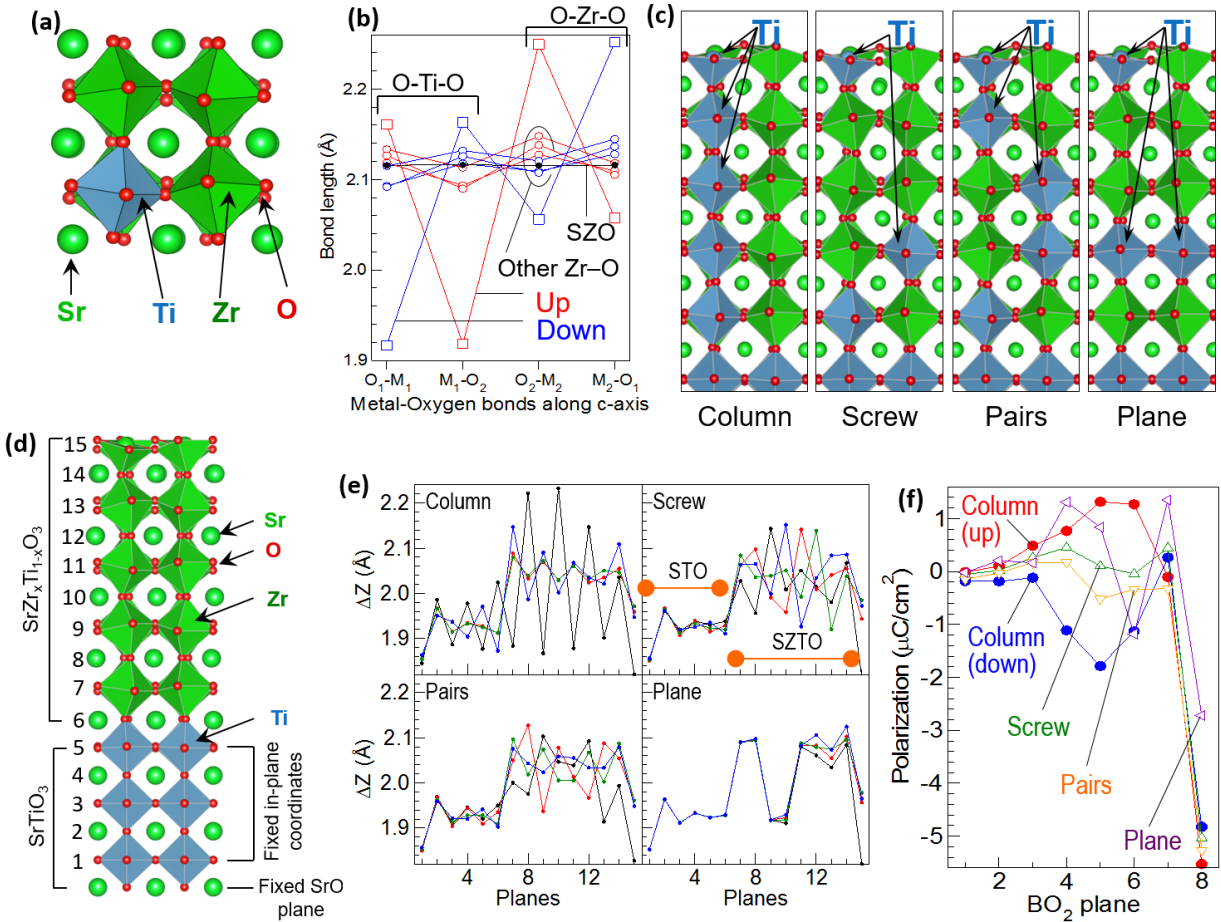

**Figure 4.** Density functional theory of SZTO. (a) Structure of the bulk SZTO with Ti concentration of 12.5% as obtained using the 2×2×2 supercell subjected to compressive strain. (b) Metal-oxygen distances in O–(Zr,Ti)–O chains along the *c*-axis show that polarization of Ti-centered octahedra induces similar polarization of Zr-centered octahedra. (c) DFT modelling of SZTO film with various configurations of Ti distributions, namely, column, screw, pairs and plane. (d) Periodic, 2×2 lateral supercell used to model SZTO film on a $SrTiO_3$ substrate (Ti dopants are not shown). Numbers on the left denote planes. (e) Separation (ΔZ) between *B*-site (Ti,Zr) cations and Oxygen anions situated in adjacent planes, stacked along the *c*-axis for the various distributions of Ti shown in (c). Colors indicate the different sets of the lateral fractional coordinates: black (0,0), red (1/2,0), green (0,1/2), blue (1/2,1/2) within the 2×2 lateral supercell. (f) The resulting out-of-plane component of polarization due to Ti and Zr cation displacements for each of the Ti configurations.